**Spin Transfer from a Ferromagnet into a Semiconductor through an Oxide barrier**

C. I. L. de Araujo[1], M. A. Tumelero[1], A. D. C. Viegas[1], N. Garcia[1,2] e A. A. Pasa[1]

[1]*LFFS, Departamento de Física, Universidade Federal de Santa Catarina, Caixa Postal 476, CEP 88040-900, Florianópolis, Santa Catarina, Brazil.*

[2]*Laboratorio de Física de Sistemas Pequeños y Nanotecnología, Consejo Superior de Investigacione, Científicas (CSIC), Serrano 144, 28006 Madrid, Spain*

Corresponding authors: andre.pasa@ufsc.br; nicolas.garcia@ufsc.br

Abstract

We present results on the magnetoresistance of the system $Ni/Al_2O_3$/n-doped $Si/Al_2O_3/Ni$ in fabricated nanostructures. The results at temperature of 14K reveal a 75% magnetoresistance that decreases in value up to approximately 30K where the effect disappears. We observe minimum resistance in the antiparallel configurations of the source and drain of Ni. As a possibility, it seems to indicate the existence of a magnetic state at the Si/oxide interface. The average spin diffusion length obtained is of 650 nm approximately. Results are compared to the window of resistances that seems to exist between the tunnel barrier resistance and two threshold resistances but the spin transfer seems to work in the range and outside the two thresholds.

Spintronic devices are based on the concept of injection and detection of spin polarized currents into non-magnetic materials. The spin diffusion length ($L_{sf}$), which is the average distance that an electron can travel in the material without a spin flip, limits the device scale of coherence. The orders of magnitude of this parameter larger in silicon than in metals [1], due to its low spin-orbit interaction, led Data and Das [2] to propose a spin-interference device based on Rashba effect [3], with spin precession controlled by a gate. The key investigation points to the experimental realization of a Field Effect Spin Transistor (SPINFET) are the measure of the $L_{sf}$ and the optimum spin polarization in the semiconductor. The solution of the drift diffusion model for current perpendicular to plane for giant magnetoresistive devices [4,5] with insertion of a spin dependent barrier in the interface between the ferromagnetic electrode and the semiconductor, which avoids the impedance mismatch problem [6,7], showed bellow,

$$r_1 = \rho_N t_N \ll r_b^* \ll \rho_N \frac{[L_{sf}]^2}{t_N} = r_2 \qquad (1)$$

establishes a condition for spin injection and detection based on the relation between interface resistance $r_b^*$, practically constant with temperature (T), and threshold resistances $r_1$, $r_2$, related to the semiconductor channel with resistivity $\rho_N$, channel length $t_N$ and spin diffusion length $L_{sf}$. Among the semiconductors studied, the most suitable is Si due to its establishment in technology. Values of $L_{sf}$ in silicon, found in literature for non-local measurements based on the Hanle effect, are very large up to 350 μm in vertical [8] and 2 mm in lateral devices [9] with injection of hot electrons in intrinsic silicon. Much lower values for $L_{sf}$, up to 300 nm, were obtained in lateral devices with diffusive injection for highly doped silicon [10]. Local measurement of spin polarized current injected and detected from ferromagnetic clusters separated in average by 50nm on low doped silicon was performed by our group with polarization up to 1% [11,12] and values up to 12% were achieved by Marukame and co-authors [13] in a nanolitographed device with a ferromagnetic/oxide/doped Si/oxide/ferromagnetic structure. However while the ref. 13 was done with an oxide barrier and at low temperature (LT), ref. 11 and 12 were done without an oxide barrier and at room temperature (RT). It is amply believed that without barrier is not possible to inject electron in the semiconductor [see ref.14]. Precisely one of the points in this paper is to discuss the value of formula (1).

In order to perform local silicon $L_{sf}$ measurement via giant magnetoresistance, nanofabricated tunnel ferromagnetic contacts on weakly doped n-type silicon substrates, with channel lengths from 100nm to 1.5 μm, are developed. The n-type Si (3 – 9 Ωcm) substrates were etched in HF diluted solution and introduced in a vacuum chamber for electron beam evaporation of a thin tunnel layer of 2 nm $Al_2O_3$, with working pressure of $3 \times 10^{-7}$ Torr. After evaporation, the Si wafer was cleaned by sonication in isopropyl alcohol (IPA) solution for 5 minutes and prebaked at 180°C on hotplate for 90 seconds. A layer with 250 nm of polimetilmetacrilate (PMMA) was spin-coated at 4000 rpm for 1 minute. A post bake of 10 minutes was carried out before the samples were inserted in the system for e-beam lithography. The electron microscopy/nanolithography work has been performed with the JSM 6490-LV / RAITH e-Line and JEM 2100F microscope(s). The photo resist was developed and followed by the deposition of 30 nm of nickel deposited by electron beam evaporation.

Now we proceed by describing our experiments. Fig.1a shows the geometry of the wires to perform the spin injection. There are three feet with one wire each one directed towards three other wires that are contacted with just one foot. There are three openings that prevent the electrical continuity of the wires. The amplified view of the opening where the electrons have to travel in the Si is also presented as an inset (right inset), while in the upper inset the geometry of the experiment is shown. The widths of the wires at each side of the contact are 140 and 360 nm. In this way, we have produced six configurations with openings of 100 to 1500 nm. Fig.1b presents the hysteresis loop of the structure with approximately 10 Oe of coercive field. We notice again that the wires are of Ni and the Si is n-doped with 6-9 ohm.cm of resistivity.

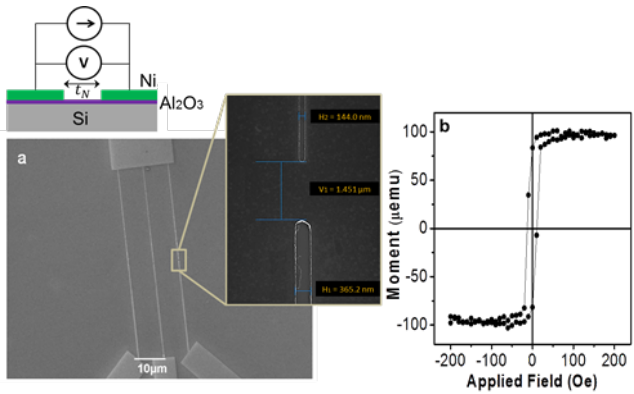

Fig.1 - (a) Structure of the wires for the experiment. From one feet come out three wires and other three wires from another three feet. The wires have a breaking that is indicated in the right inset. The upper inset shows the geometry of the experiment. Fig.1b, hysteresis curve with 10 Oe coercive field.

Experiments are presented in Figures 2 to 4. In the magnetoresistance (MR) curve displayed in Fig. 2, for in-plane measurements with the field transversal to the current, an effect reaching up 75% at 14 K is observed. We should say that this is for a current of 100 µA, however the values of the MR keep practically constant for all the currents we have tried from 10 to 500 µA. Although we find that the MR curves are more open and less saturated at low currents. This is probably a consequence of the non-linear resistance due to the existing tunnel and Schottky barriers. In addition to high values of MR observed, we have obtained a surprising result and this is the minimum in the resistance for low magnetic fields. Since the applied current is constant, this implies that the voltage is a minimum in the antiferromagnetic configuration of source and drain. But this cannot be understood because the two electrodes are of Ni. The only way we find an explanation for these measurements is that a magnetic state is present at the interface $Al_2O_3$/Si, as we already observed in ref. 12. The right hand inset is an amplification of the data at low T showing a displacement of the curves of about 10 Oe, which is approximately the coercive field measured in Fig. 1. The left-handed inset is a plot of the $r_1$, $r_2$ and $r_b^*$. It should be noticed that up 20 K the $r_b^*$ is between the two resistance window expressed in Eq. (1), but in the temperature interval of 20 to 30 K the $r_b^*$ is above the upper limit of the window. At this point we should add the information that when the field is applied in plane in the direction of the current (longitudinal configuration) the MR value measured is about 30% and has a wider peak than the transversal case.

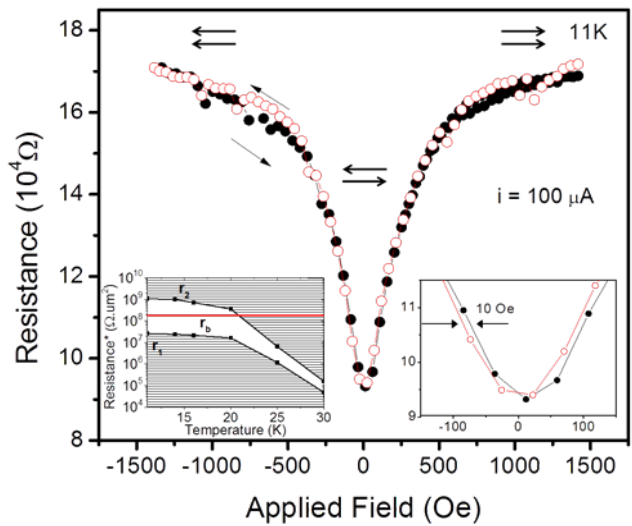

Fig.2- Resistance curve with a minimum for the antiferromagnetic configuration of source and drain for in-plane measurements with the magnetic field applied transversal to the electric current. Right inset is amplification near to minimum to show that there is 10Oe coercive field as measured in Fig.1b. Right inset is the plot of the two limiting resistances, formula (1), for 100 nm channel length, as a function of T and the horizontal line is the $r_b^*$. It can be seen that up to 20K formula (1) is satisfied but from 20 to 30K does not work.

A resume of the obtained data is given in Fig. 3. In Fig. 3a are MR data as a function of the magnetic field and temperature T(K). We can see that at 30 K the MR is very small around 1%. Fig. 3b represents the MR curves, at 14 K, as a function of the magnetic field and the opening distances ($t_N$). The MR decreases with increasing values of distance between the contacts, due to successive spin collisions. Notice that the spin accumulation varies as $\exp(-L_{sf}/t_N)$. So these curves express that the MR reduces with increasing T and the length of the opening or Si channel.

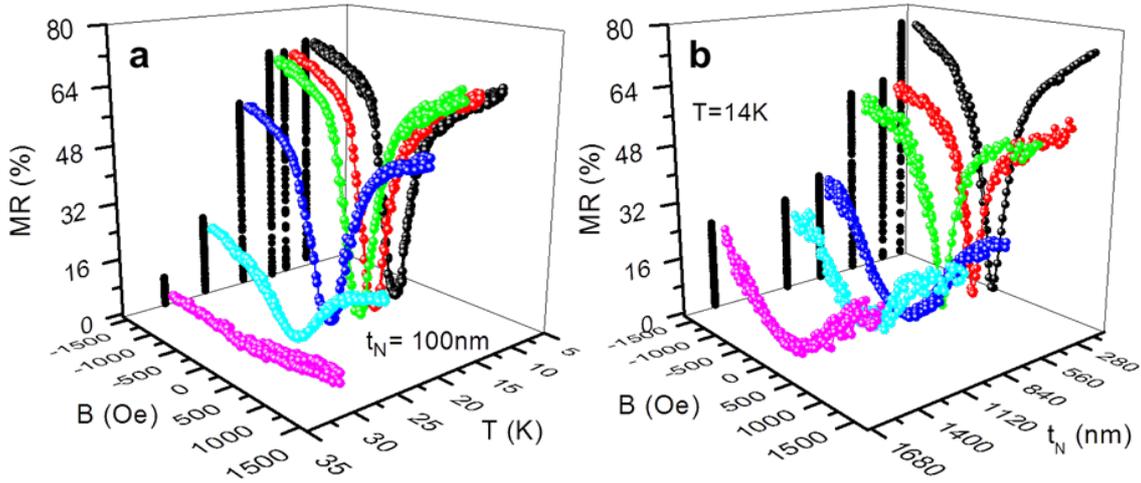

Fig.3- (a) MR as a function of T and of the applied magnetic field. MR vanishes drastically at 30K. (b) The same but now at T=14K and as a function of the channel length.

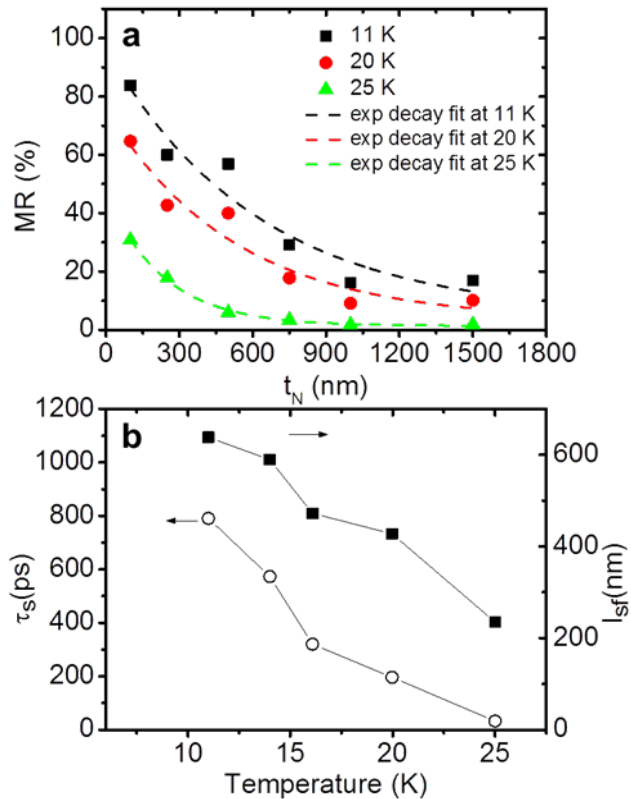

Fig.4- a) MR at different T versus $t_N$, by fitting this curves with an exponential term (see text) one obtain the values of spin diffusion length. b) Spin diffusion length and spin relaxation time as a function of temperature.

From the data of MR vs. $t_N$ we can obtain the $L_{sf}$ or spin diffusion length by fitting a exponential term exp(-$L_{fs}$/$t_N$) for each T as in Fig. 4a for the temperatures of 11, 20 and 25 K. The values of $L_{fs}$ versus T are depicted in Fig. 4b together with the spin relaxation time $\tau_s$ that can be obtained from the formula, $L_{sf} = \sqrt{D\tau}$, where D is diffusion coefficient. The value of $L_{fs}$ at 14 K is 650 nm for example. The value of 200 ps for $\tau_s$ at 20 K is comparable to the one obtained in ref. 10.

From our experiments we have found the spin transfer in a sandwich formed by a Ni/oxide/doped Si/oxide/Ni and that the voltage is maximum for constant applied currents in the ferromagnetic configuration of source and drain and a large value of MR=75% has been obtained for T = 14 K. A similar result is also obtained in ref. 13 where a similar result has been reported but with a 15% MR using a Heusler ferromagnetic at 10 K. As we say above, this is surprising and difficult to understand because the minimum of Fig. 2 should be a maximum. It could be understood if a magnetic state exists at the silicon interface with magnetization and majority density of carries at the Fermi level in the same direction because Ni have them in opposite direction. In addition, the window of resistance of formula (1) is satisfied up to 20 K but after that it fails drastically as can be observed from the right hand side inset of Fig.2. More experiments are needed to validate this point using structures with and without tunnel barriers, for example.

Acknowledgments:

This work was financially supported by Brazilian (CNPQ, CAPES, FINEP and FAPESC) and Spanish (CSIC) agencies. The electron nanolithography work was performed at LABNANO/CBPF, Rio de Janeiro, Brazil, with strong support for the fabrication of the samples from Dr. H. Lozano and Prof. L. Sampaio. The SEM images were obtained at LABNANO/CBPF and LCME/UFSC.